\documentclass{LT23auth}
\usepackage{graphicx}

\begin{document}

\begin{frontmatter}

\title{Spin collective mode and quasiparticle contributions to STM spectra of
$d$-wave superconductors with pinning}

\author[address1]{Anatoli Polkovnikov},
\author[address1]{Subir Sachdev \thanksref{thank1}}, and
\author[address2]{Matthias Vojta}

\address[address1]{Department
of Physics, Yale University, P.O. Box 208120, New Haven CT
06520-8120, USA}

\address[address2]{Theoretische Physik III, Elektronische Korrelationen
und Magnetismus, Institut f\"ur Physik,\\ Universit\"at Augsburg,
D-86135 Augsburg, Germany}

\thanks[thank1]{E-mail: subir.sachdev@yale.edu; Fax: +1-203-432-6175}

\begin{abstract}
We present additional details of the scanning tunneling microscopy
(STM) spectra predicted by the model of pinning of dynamic spin
collective modes in $d$-wave superconductor proposed by
Polkovnikov {\em et al.} (Phys. Rev. B {\bf 65}, 220509 (2002)).
Along with modulations at the twice the wavevector of the spin
collective mode, the local density of states (LDOS) displays
features linked to the spectrum of the Bogoliubov quasiparticles.
The former is expected to depend more strongly on an applied
magnetic field or the doping concentration. The spin collective
mode and the quasiparticles are distinct, co-existing, low energy
excitations of the $d$-wave superconductor (strongly coupled only
in some sectors of the Brillouin zone), and should not be viewed
as mutually exclusive sources of LDOS modulation.
\end{abstract}

%
%
\end{frontmatter}

\section{Introduction}

The observation of a halo of checkerboard modulation in the local
density of states (LDOS) around magnetic field-induced vortex
cores by Hoffman {\em et al.} \cite{seamus} in optimally doped
Bi$_2$Sr$_2$CaCu$_2$O$_{8+\delta}$ has focused attention on the
nature of the spin and charge collective modes in the cuprate
superconductors. Charge order in this region was anticipated
theoretically \cite{kwon,dsz,vs99}. By `charge order' we mean here
periodic spatial modulations of all local observables invariant
under spin rotations and time-reversal, such as the exchange and
pairing energies, and the LDOS itself. If the modulation is
primarily on quantities defined on a bond, rather than in the
charge density on a site, then such order is better denoted as
`bond order'; it was this bond order which was theoretically
anticipated \cite{kwon,vs99}. The general connection relating such
charge/bond order to the `square' of collinear, non-bipartite,
spin correlations \cite{zachar,zaanen,schulz,machida} (see
Section~\ref{model} below) was used to develop a model
\cite{psvd,zds,pvs} relating the spatial structure and energy
spectrum of the charge order to the {\em dynamic} spin fluctuation
spectrum as measured by neutron scattering
\cite{katano,lake,boris,lake2}. There have also been other
discussions of competing order within vortex cores, including
enhanced dynamic antiferromagnetism in a spin-gap state
\cite{ssvortex,nl}, static N\'{e}el order \cite{arovas}, and other
models with static spin density wave order
\cite{chenting,zhu,franz,zhang,ghosal,andersen02}.

More recent measurements of Howald {\em et al.} \cite{aharon} have
shown that the LDOS modulations in the vortex halo also persists
in zero magnetic field, albeit with a much weaker amplitude: here
the pinning of the charge/bond order is presumably due to
impurities and other imperfections in the
Bi$_2$Sr$_2$CaCu$_2$O$_{8+\delta}$ crystal. Theoretical modeling
of the STM spectrum has indicated \cite{podolsky,matthias} that
the order is likely to be primarily on bond variables, as may be
derived from a spin-Peierls state \cite{sns}, {\em i.e.\/} a {\em
bond order wave} \cite{campbell}. This is in contrast to the
prediction of a charge density modulation on the Cu sites in the
anti-phase domain wall picture of `stripes'
\cite{zaanen,schulz,machida}, which is not observed \cite{aharon}.

Hoffman {\em et al.} \cite{seamus2} have also reported related
measurements in zero magnetic field, and have focused attention on
modulations of the LDOS at wavevectors throughout the Brillouin
zone, and on their dispersion as a function of energy. They have
identified some wavevectors with quasiparticle excitations of the
superconductor scattered by random impurities, as suggested in
Refs.\cite{byers,dunghai}.

The purpose of the present paper is to discuss the relationship
between the quasiparticle and collective mode contributions to the
LDOS modulations in a $d$-wave superconductor. We will use
precisely the model proposed earlier by Polkovnikov {\em et al.}
\cite{pvs} for the pinning of the spin collective mode: we show
that the earlier expressions for the induced LDOS modulations also
contain additional contributions which can be interpreted as those
due to quasiparticle scattering, with a spectrum related to
observations \cite{seamus2}. The relative strength of the
collective and quasiparticle contributions can be varied by
changing the gap to the spin collective mode: experimentally the
gap can be reduced (and the collective mode contribution enhanced)
by lowering the doping or by applying a magnetic field \cite{dsz},
and this should allow a test of our predictions. Some related
observations have also been made independently by Han \cite{han}.
Our main conclusion is that the LDOS modulations observed by STM
contain contributions from both the spin collective mode and the
quasiparticles: as described in our simple model \cite{pvs}, these
are relatively independent low energy excitations of the BCS
superconductor, but they can become strongly coupled in some
sectors of the Brillouin zone if permitted by energy and
wavevector conservation \cite{did}; they should not be viewed as
mutually exclusive sources of LDOS modulations.

\section{Model}
\label{model}
We assume the cuprate superconductor has a dominant
spin collective mode near the wavevectors ${\bf K}_x=(3\pi/4,\pi)$
and ${\bf K}_y=(\pi,3\pi/4)$, and so write for the spin operator
on the lattice site ${\bf r}$ at imaginary time $\tau$
\begin{equation}
S_{\alpha} ({\bf r}, \tau) = \mbox{Re} \left[e^{i {\bf K}_{x}
\cdot {\bf r}} \Phi_{x \alpha} ({\bf r}, \tau)+  e^{i {\bf K}_{y}
\cdot {\bf r}} \Phi_{y \alpha}({\bf r}, \tau)\right], \label{eq1}
\end{equation}
where $\Phi_{x,y \alpha}$ are the spin density wave (SDW) order
parameters which are assumed to be smooth functions of spacetime.
As an aside, and for completeness, we mention that charge/bond
order parameters can also be constructed out of the $\Phi_{x,y
\alpha}$. We define
\begin{eqnarray}
Q_{{\bf a}} ({\bf r}, \tau) &\equiv& S_{\alpha} ({\bf r}, \tau)
S_{\alpha} ({\bf r} + {\bf a}, \tau) \nonumber \\
&\approx & \mbox{Re} \left[ e^{2 i {\bf K}_{x} \cdot {\bf r} + i
{\bf K}_x \cdot {\bf a}} \Phi_{x \alpha}^2 ({\bf r}, \tau) \right]
+ \ldots \label{eq2}
\end{eqnarray}
where ${\bf a}$ is a vector representing a bond (say the nearest
neighbor vector), we have assumed that $\Phi_{x\alpha}$ does not
vary significantly over the spatial distance ${\bf a}$, and the
ellipses denote numerous other similar terms which can be deduced
from (\ref{eq1}). For ${\bf a}$ a nearest-neighbor vector,
$Q_{{\bf a}}$ measures the modulations in the exchange energy,
while for ${\bf a} = 0$, $Q_{{\bf a}}$ measures the local charge
density (for the $t$-$J$ model the charge density is linearly
related to the on-site $S_{\alpha}^2$). Thus we see from
(\ref{eq2}) that, as noted earlier, the charge/bond order
parameter is proportional to the square of the SDW order
parameter. In a similar manner, we will see below that the LDOS
modulations are also sensitive to the square of the SDW order and
so occur at wavevectors $2 {\bf K}_{x,y}$: however our computation
will include retardation effects missing from a phenomenological
correspondence analogous to (\ref{eq2}). All our computations
below will be in a regime where
\begin{equation}
\langle S_{\alpha} \rangle = \langle \Phi_{x \alpha} \rangle =
\langle \Phi_{y \alpha} \rangle = 0
\end{equation}
(because spin rotation invariance is preserved), but we will
nevertheless have the rotationally invariant
\begin{equation}
\sum_{\alpha} \langle \Phi_{x,y \alpha}^2 \rangle \neq 0,
\end{equation}
because translational invariance is broken by a pinning
potential. This should be contrasted with the distinct physical
pictures of
Refs.~\cite{arovas,chenting,zhu,franz,zhang,ghosal,andersen02} all
of which break spin rotation invariance with $\langle S_{\alpha}
\rangle \neq 0$.

To keep this paper self-contained, we now briefly describe the
model used to compute the LDOS modulations; for details see
Ref.~\cite{pvs}. The effective action for the $\Phi_{x,y\alpha}$
fluctuations is assumed to be Gaussian:
\begin{equation}
\mathcal{S}_{\Phi} = T \sum_{{\bf q}, \omega_n, \alpha}
\chi^{-1}_x ( {\bf q}, \omega_n ) | \Phi_{x \alpha} ({\bf q},
\omega_n) |^2 + (x \rightarrow y)
\end{equation}
where $\omega_n$ are Matsubara frequencies. The $\chi_{x,y} ({\bf
q},\omega)$ denote the dynamic spin susceptibility as a function
of wavevector ${\bf q}$ (measured from the ordering wavevectors
${\bf K}_{x,y}$) and frequency $\omega$, which is measured by
inelastic neutron scattering. In the absence of available
experimental input for Bi$_2$Sr$_2$CaCu$_2$O$_{8+\delta}$, we will
continue to use the simple phenomenological form \cite{pvs}
expected near a SDW ordering quantum phase transition in a
superconductor\cite{zds}: $\chi^{-1} ( {\bf q}, \omega_n
)=\omega_n^2+c^2 {\bf q}^2+\Delta^2$, where $c$ is a spin wave
velocity, and $\Delta$ is the spin gap to collective excitations
near the wavevectors ${\bf K}_{x,y}$ (we have assumed $\chi_x =
\chi_y \equiv \chi$ for simplicity). Han \cite{han} has used the
same model, but with microscopic computations on the lattice
$t$-$J$ model as input for $\chi$.

The all-important ingredient which produces an observable {\em static}
modulation in the STM signal is the pinning of the sliding degree
of freedom of the SDW fluctuations.
This pinning is realized by any imperfection (impurity, vortex core, etc.)
which breaks the lattice translational symmetry.
We employ the simplest possible pinning term which is
invariant under the SU(2) spin symmetry,
\begin{equation}
\mathcal{S}_{\rm pin} = -\sum_{\alpha} \int d \tau \left[ \zeta_x
\Phi_{x \alpha}^2 ({\bf r}_0, \tau) + (x\rightarrow y) + c.c.
\right], \label{spin}
\end{equation}
where $\zeta_{x,y}$ are complex coupling constants representing
the pinning potential.

The electronic quasiparticles of the $d$-wave superconductor
are described by a standard BCS model:
\begin{equation}
H_{\rm BCS} = \sum_{\bf k} \Psi^{\dagger}_{\bf k} \left[
(\varepsilon_{\bf k} - \mu) \tau^z + \Delta_{\bf k} \tau^x \right]
\Psi_{\bf k}. \label{hbcs}
\end{equation}
Here $\Psi_{\bf k} = (c_{{\bf k}\uparrow}, c_{-{\bf k}\downarrow}^{\dagger})$ is
a Nambu spinor at momentum ${\bf k}=(k_x, k_y)$,
$\tau^{x,y,z}$ are Pauli matrices in particle-hole space, and $\mu$ is
a chemical potential.
For the kinetic energy, $\varepsilon_{\bf k}$ we
have first ($t$) and second ($t^{\prime}$) neighbor hopping,
and the gap function is $\Delta_{\bf k} = (\Delta_0/2)
(\cos k_x - \cos k_y )$.
The full Hamiltonian for the conduction electrons
includes a coupling to the collective SDW fluctuations:
\begin{eqnarray}
H &=& H_{BCS} + \gamma \sum_{{\bf r}} c^{\dagger}_{\mu} ({\bf r})
\sigma^{\alpha}_{\mu\nu} c_{\nu} ({\bf r}) \times \\
&\times& \left( \Phi_{x \alpha} ({\bf r}) e^{i {\bf K}_x \cdot
{\bf r}} + \Phi_{x \alpha}^{\ast} ({\bf r}) e^{-i {\bf K}_x \cdot
{\bf r}} + (x\rightarrow y)  \right), \nonumber
\end{eqnarray}
where $\sigma^{\alpha}$ are the Pauli spin matrices, and $\gamma$
describes the scattering of the Bogoliubov quasiparticles of the
superconductor off the SDW fluctuations.

The predictions of this model \cite{pvs} for the quasiparticle
LDOS, to second order in the quasiparticle-collective mode
coupling $\gamma$ and to first order in the pinning strength
$\zeta$, can be written as
\begin{eqnarray}
&&\delta \rho({\bf r},\omega) \approx {\rm Im}\,{\rm Tr}{1+\tau_z\over 2}\times\nonumber \\
&&\phantom{X} \sum_{{\bf r}_1, {\bf r }_2} G_0({\bf r}-{\bf
r}_1,\omega)\Sigma({\bf r}_1,{\bf r}_2,\omega) G_0({\bf r}_2-{\bf
r},\omega), \label{e1}
\end{eqnarray}
where $G_0$ is the electronic Green's function in the canonical
translationally invariant BCS superconductor described by
(\ref{hbcs}). There is a contribution to the self energy
$\Sigma=\Sigma_{\rm sp}$ from the coupling to the spin collective
mode, and we focus only on the part which breaks translational
invariance because of the pinning of the collective mode
\begin{eqnarray}
&&\Sigma_{\rm sp} ({\bf r}_1,{\bf r}_2,\omega) \approx 12\zeta
\gamma^2 \int
{d\nu\over 2\pi} \chi({\bf r}_2,\nu) \chi({\bf r}_1,\nu) \nonumber \\
&&~~~~\times G_0({\bf r}_2-{\bf r}_1,\omega-\nu) \cos ({\bf K}_x
\cdot ({\bf r}_1+{\bf r}_2) -\delta) \nonumber \\
&&~~~~+(x \rightarrow y). \label{e2}
\end{eqnarray}
The parameter $\delta$, which is the phase of the coupling
$\zeta$, distinguishes site centered SDWs ($\delta=0$) and bond
centered SDWs ($\delta=\pi/2$). Following the STM
experiments~\cite{seamus,aharon,seamus2}, we will compute the
Fourier transform of the LDOS:
\begin{eqnarray}
&&\delta\rho_{\rm sp} ({\bf q},\omega)=\nonumber{\rm
Tr}{1+\tau_z\over 4} \, \big\{\mbox{Im} \int {d^2 k\over(2\pi)^2}
\, G_0({\bf k },\omega) \times \\ && [ \Sigma_{\rm sp} ({\bf
k},{\bf k}+{\bf q},\omega) + \Sigma_{\rm sp} (-{\bf k},-{\bf
k}-{\bf q},\omega) ] G_0({\bf k}+{\bf q} ,\omega) \nonumber \\ &&
- i\, \mbox{Re} \int {d^2 k\over(2\pi)^2} \, G_0({\bf k },\omega)
[ \Sigma_{\rm sp} ({\bf k},{\bf k}+{\bf q},\omega) - \nonumber\\
&& - \Sigma_{\rm sp} (-{\bf k},-{\bf k}-{\bf q},\omega) ] G_0({\bf
k}+{\bf q} ,\omega) \big\}\,. \label{e3}
\end{eqnarray}
Note that the imaginary part of this expression vanishes in the
presence of inversion symmetry,
$\delta \rho({\bf r},\omega) = \delta \rho(-{\bf r},\omega)$.
The results (\ref{e2},\ref{e3}) are for a single pinning center,
and the realistic many impurity case will be similar provided
interference between impurities is small.

It is not difficult to see from (\ref{e1}-\ref{e3}) that in the
limit of a large spin gap ($\Delta \rightarrow \infty$), the
modulations in the LDOS predicted by (\ref{e3}) are identical to
those predicted in a model of quasiparticle scattering off
isolated, point-like static impurities or other imperfections
considered in Ref.~\cite{dunghai}. For a general scattering
potential $u({\bf q})$, and impurity-induced variation in the
superconducting gap $v({\bf q}_1,{\bf q}_2)$, the expression for
the LDOS modulation due to {\em static} impurities in the Born
approximation, $\delta\rho_{\rm imp} ({\bf q},\omega)$, is
formally identical to Eq. (\ref{e3}) with $\Sigma_{\rm sp}$ replaced by
\begin{equation}
\Sigma_{\rm imp}({\bf k},{\bf k}+{\bf q},\omega) = u({\bf q})\tau_z+v({\bf
k},{\bf k+q})\tau_x \,.
\label{qs}
\end{equation}
The following section will compare the predictions of (\ref{qs})
to those in (\ref{e3}) of a pinned dynamic SDW with a finite
$\Delta$.

It is interesting to note from (\ref{e3},\ref{qs}) that for any
real scattering potential, with $u^{\ast} ({\bf q}) = u(-{\bf
q})$, the LDOS modulation at wavevector ${\bf q}$ is determined by
$\int d^2 k \mbox{Im}  \left[ G_0 ({\bf k}) G_0 ({\bf k} + {\bf
q}, \omega) \right]$. Contributions proportional to $\int d^2 k
\mbox{Im} \left[ G_0 ({\bf k})\right] \mbox{Im} \left[ G_0 ({\bf
k} + {\bf q}, \omega) \right]$, which are used by
Ref.~\cite{dunghai} to explain the STM observations of
Refs.~\cite{seamus2,seamus3}, can only come from lifetime effects
which produce a complex self energy.

\section{Results}

\begin{figure}[ht]
\begin{center}\leavevmode
\includegraphics[width=0.72\linewidth]{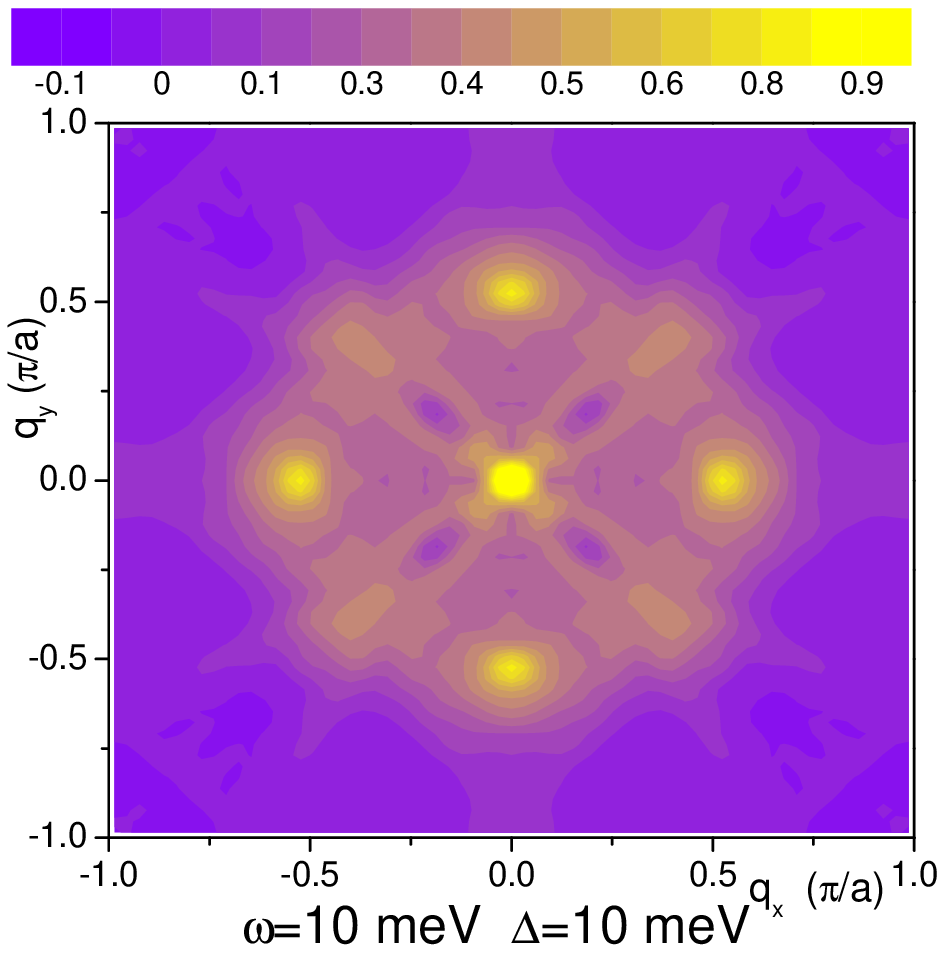}
\includegraphics[width=0.72\linewidth]{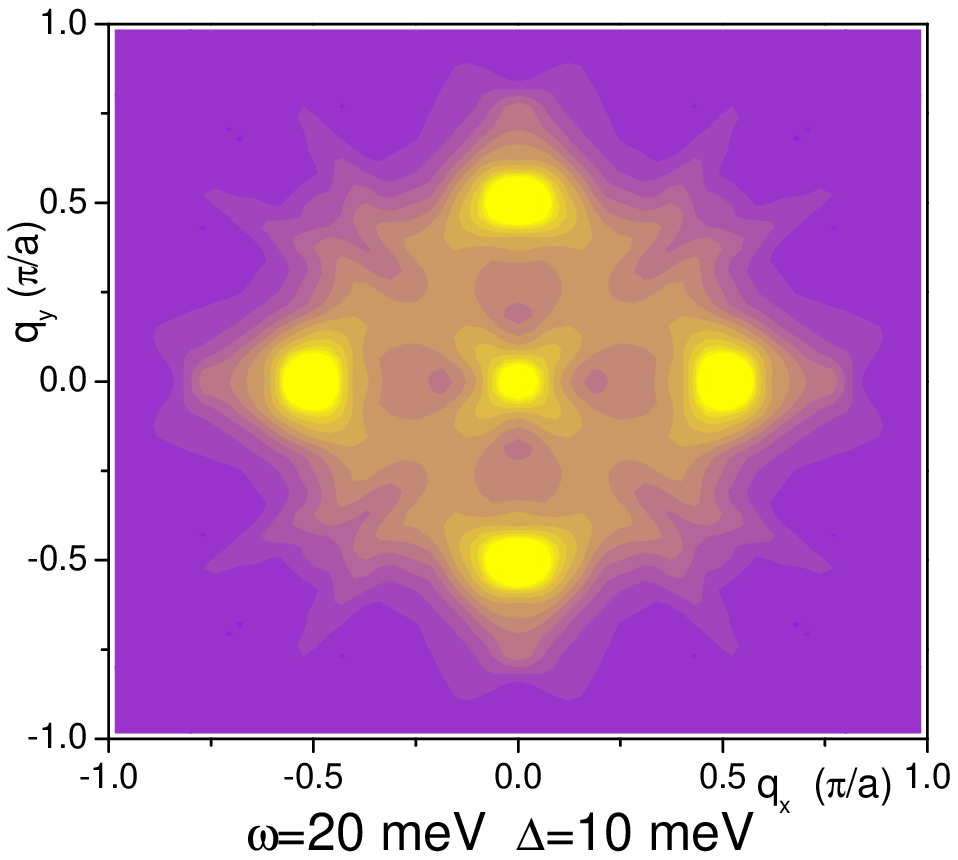}
\includegraphics[width=0.72\linewidth]{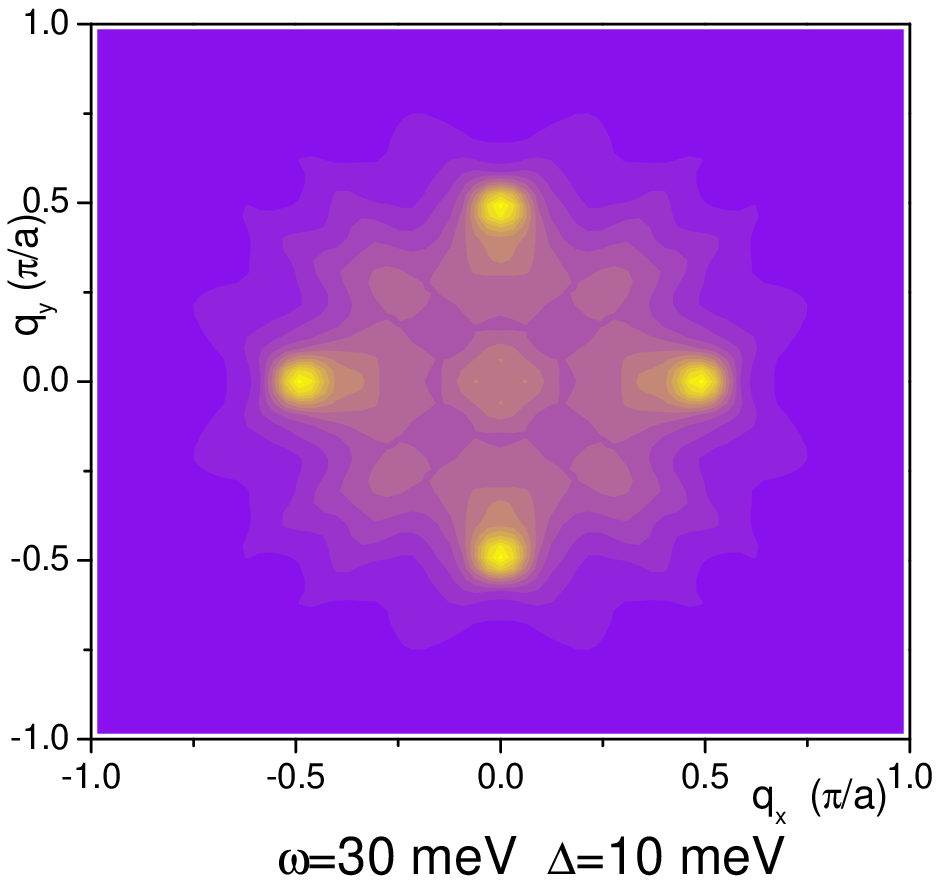}
\caption{Fourier map of the LDOS $\delta \rho_{\rm sp}({\bf q},
\omega)$ in (\protect\ref{e3}) for $\Delta=10$ meV and
quasiparticle dispersion $\varepsilon_{\bf k}$ and pairing
amplitude $\Delta_{\bf k}$ chosen as in Ref.~\cite{pvs}. The
maxima around $(\pi/2,0)$ come mainly from scattering off a pinned
SDW. The other peaks are due to details of quasiparticle
spectrum.} \label{Fig. 1}\end{center}
\end{figure}

Sample results from the dynamic SDW model (\ref{e3}) are shown in
Figs~\ref{Fig. 1}-\ref{Fig. 3}.
\begin{figure}[ht]
\begin{center}\leavevmode
\includegraphics[width=0.72\linewidth]{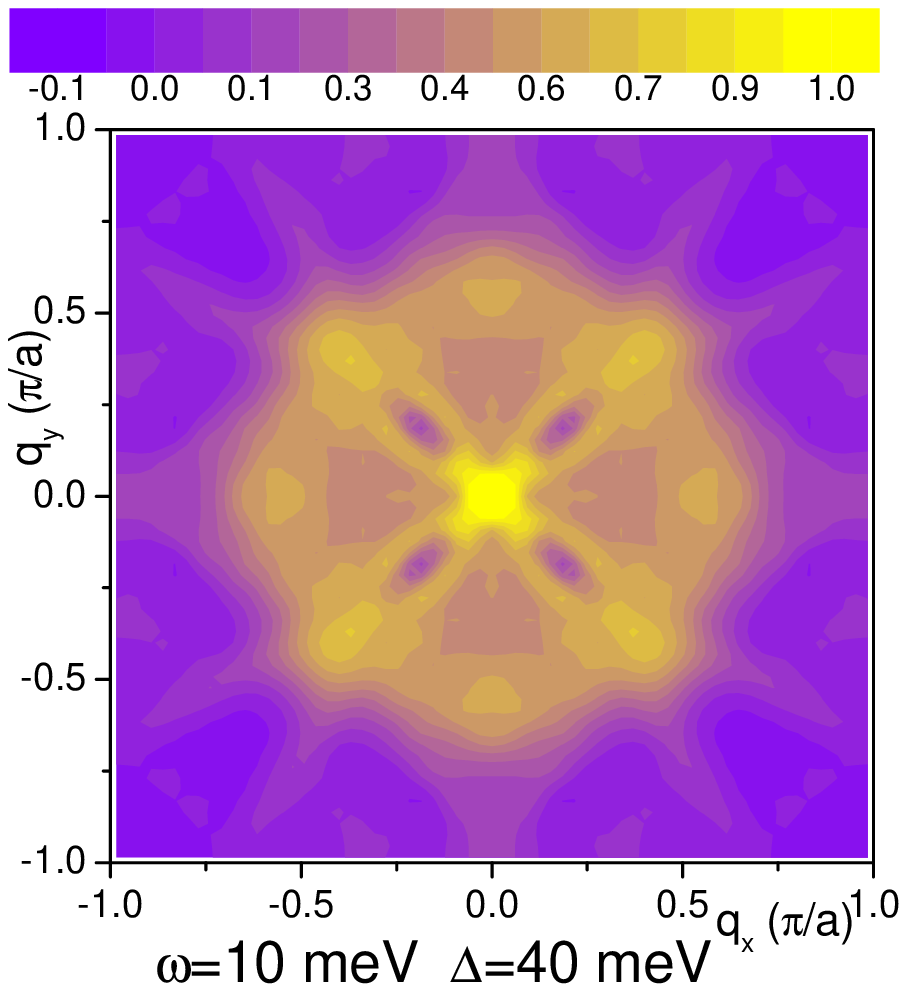}
\includegraphics[width=0.72\linewidth]{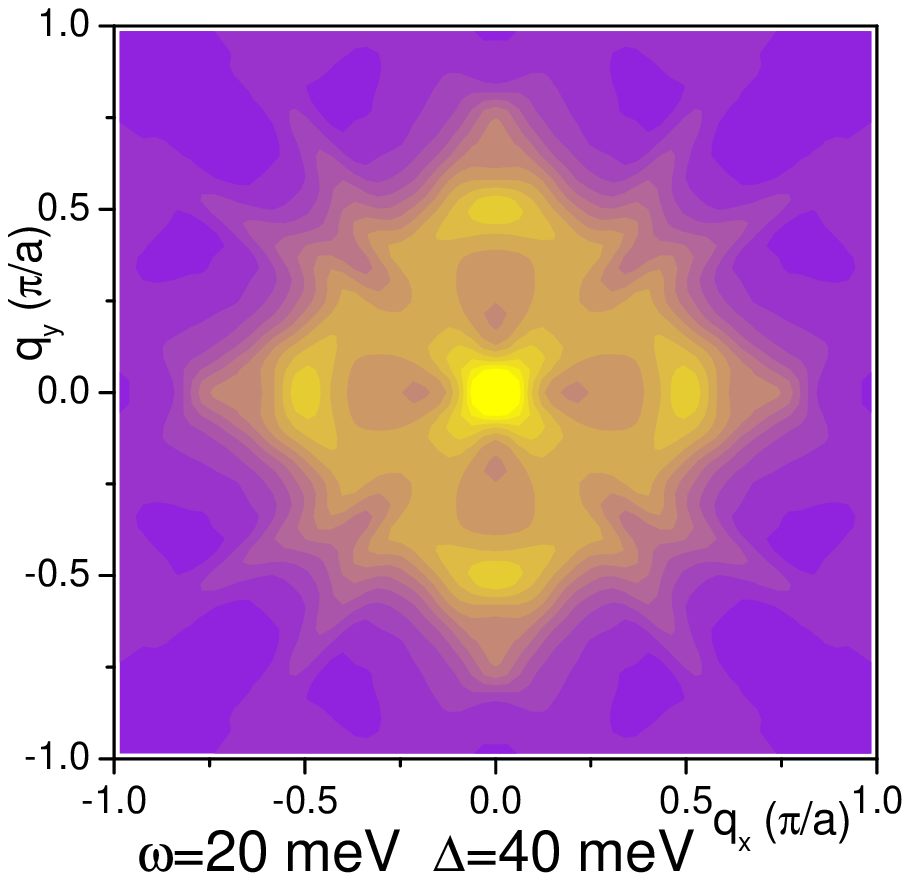}
\includegraphics[width=0.72\linewidth]{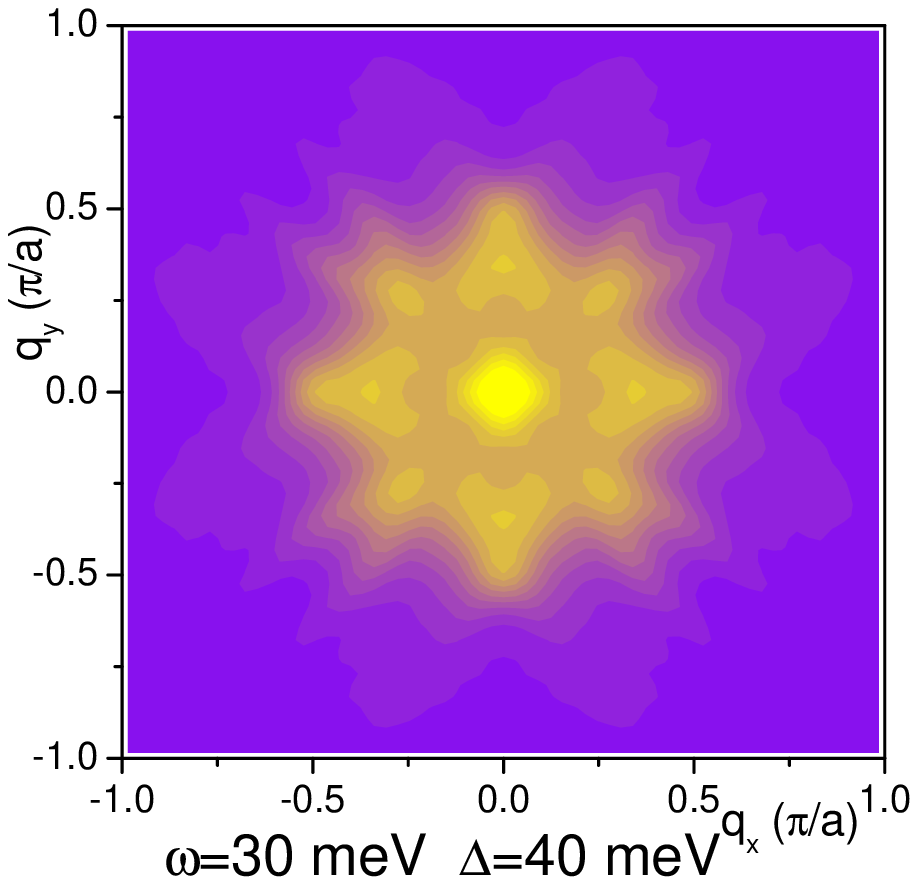}
\caption{As in fig.~\ref{Fig. 1} for $\Delta=40$ meV.} \label{Fig.
2}\end{center}
\end{figure}
For small spin gap $\Delta$, the LDOS shows the expected
\cite{zachar,psvd} modulations near ${\bf q}=2 {\bf K}_{x,y}$. The
magnitude of this modulation depends strongly on the value of
$\Delta$, and the strong expected dependence of $\Delta$ on doping
and applied magnetic field \cite{dsz} ($\Delta$ decreases with
decreasing $\delta$ and increasing field) will lead to a
corresponding variation in the LDOS peaks near ${\bf q} = 2 {\bf
K}_{x,y}$. This peak also disperses as a function of $\omega$,
with the dispersion being stronger for larger values of $\Delta$
as shown in Fig.~\ref{Fig. 3}; the decrease in the ${\bf q}$ value
of the peak with increasing $\omega$ is consistent with
observations \cite{seamus2}.
\begin{figure}[ht]
\begin{center}\leavevmode
\includegraphics[width=0.66\linewidth]{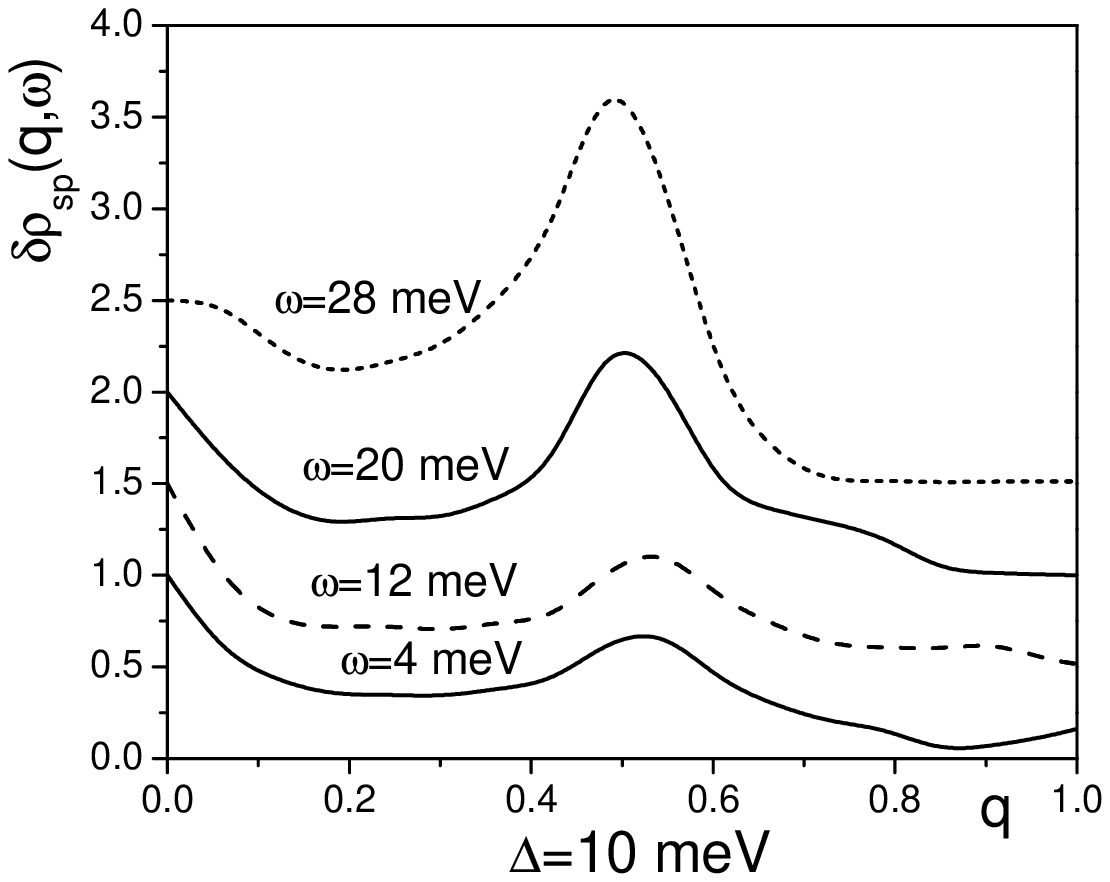}
\includegraphics[width=0.66\linewidth]{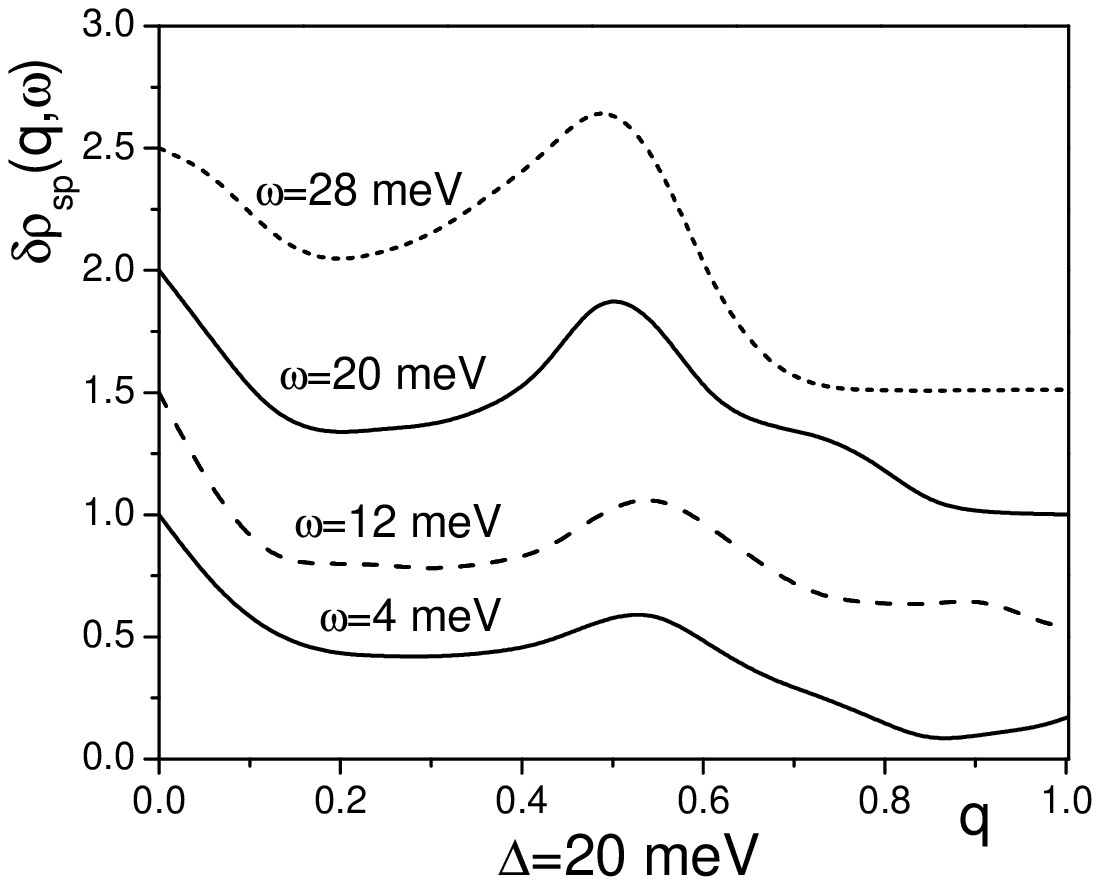}
\includegraphics[width=0.66\linewidth]{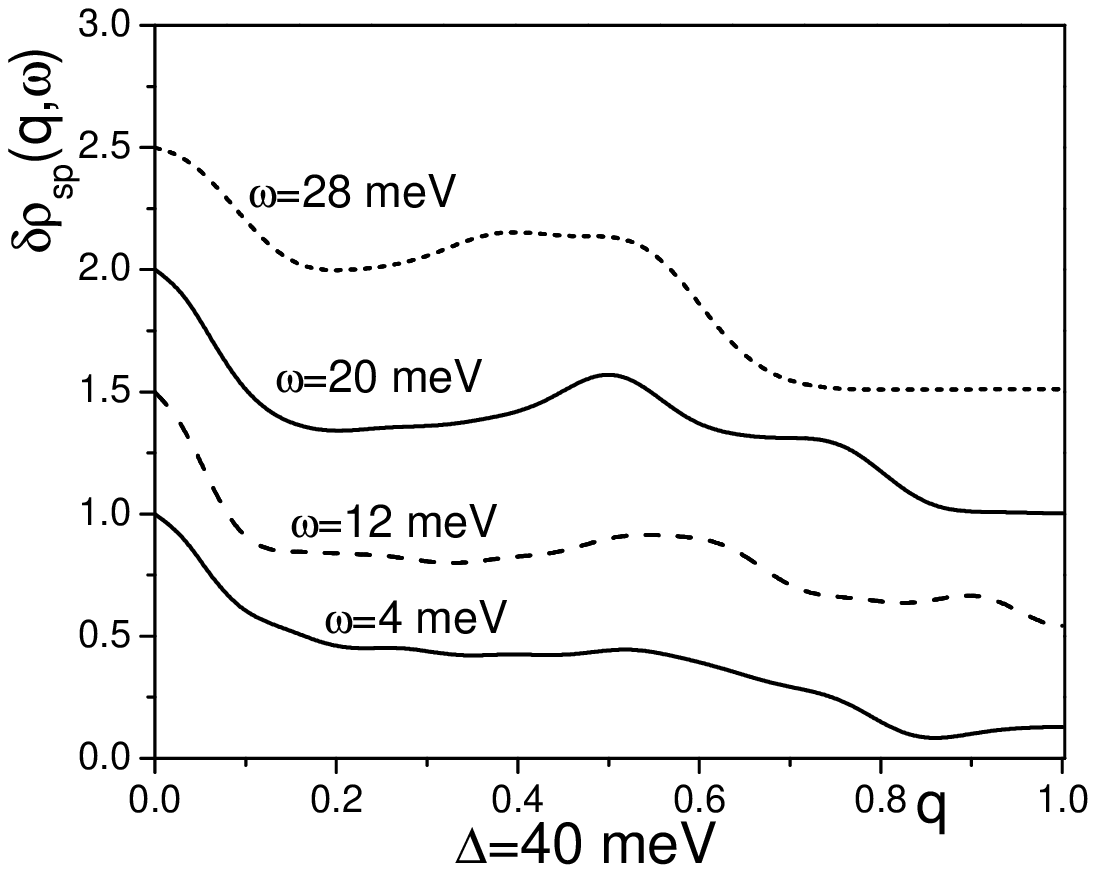}
\caption{Fourier component of the LDOS, $\delta \rho_{\rm sp}
({\bf q},0)$, for ${\bf q} = (q\pi/a,0)$; a large contribution
near $q=0$ is not shown.} \label{Fig. 3}\end{center}\end{figure}

\begin{figure}[ht]
\begin{center}\leavevmode
\includegraphics[width=0.72\linewidth]{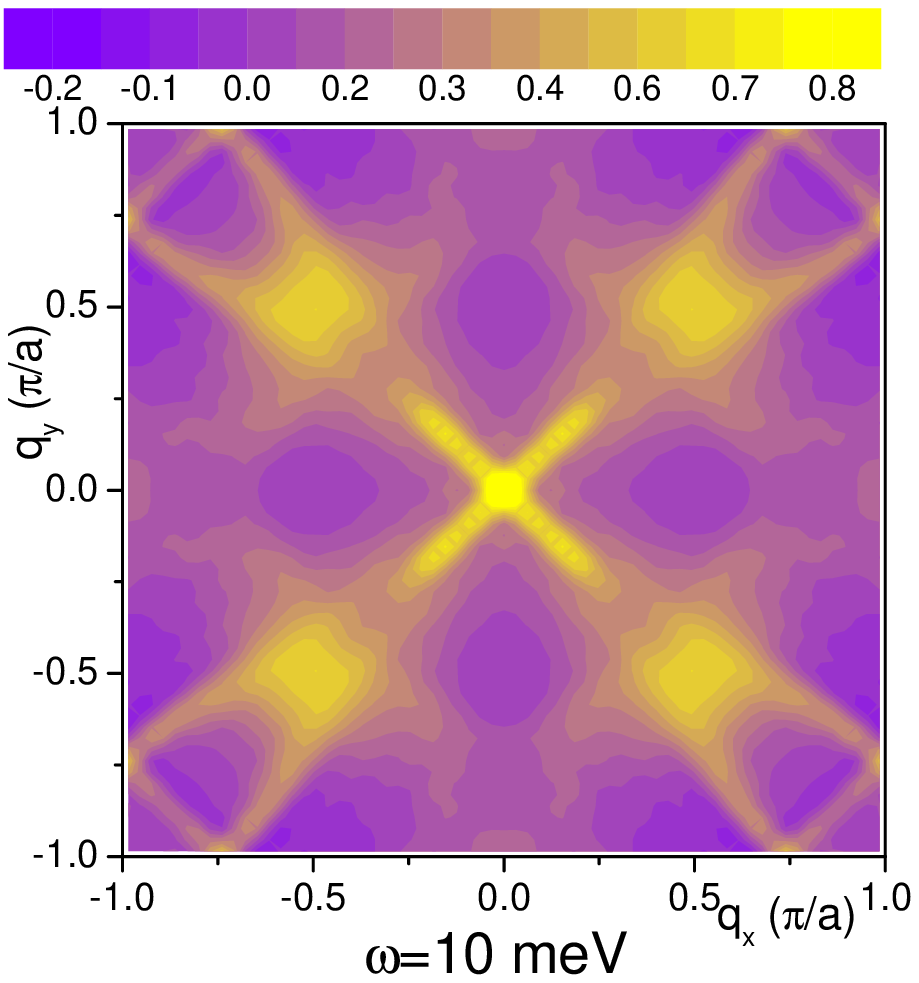}
\includegraphics[width=0.72\linewidth]{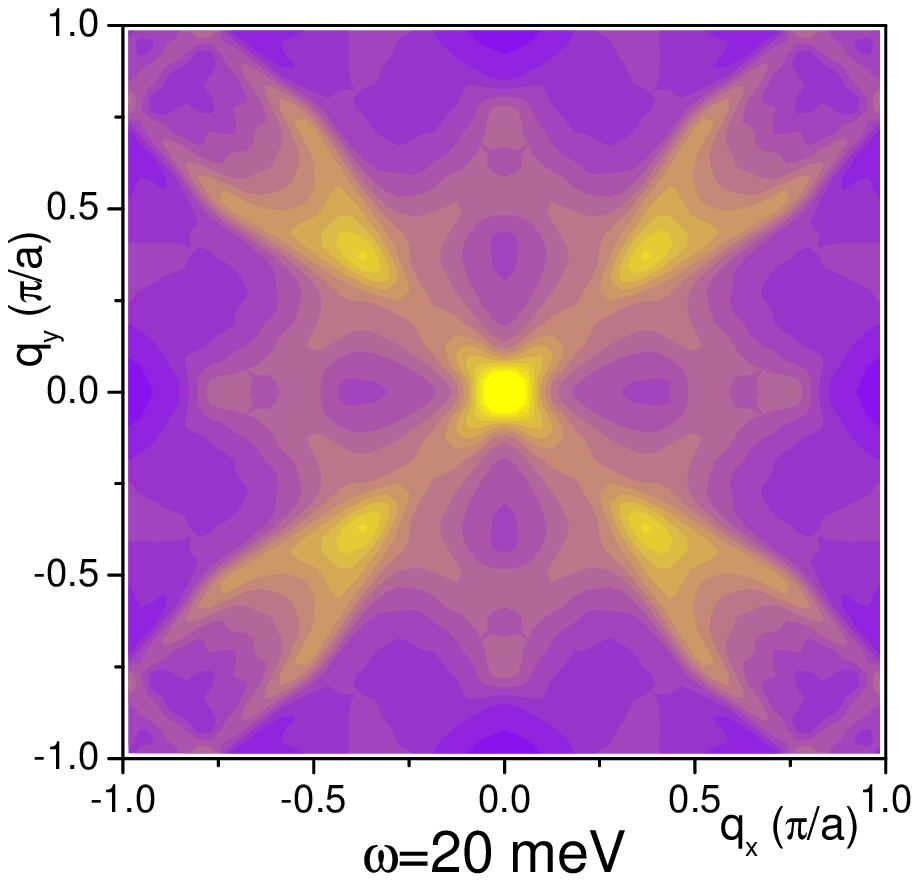}
\includegraphics[width=0.72\linewidth]{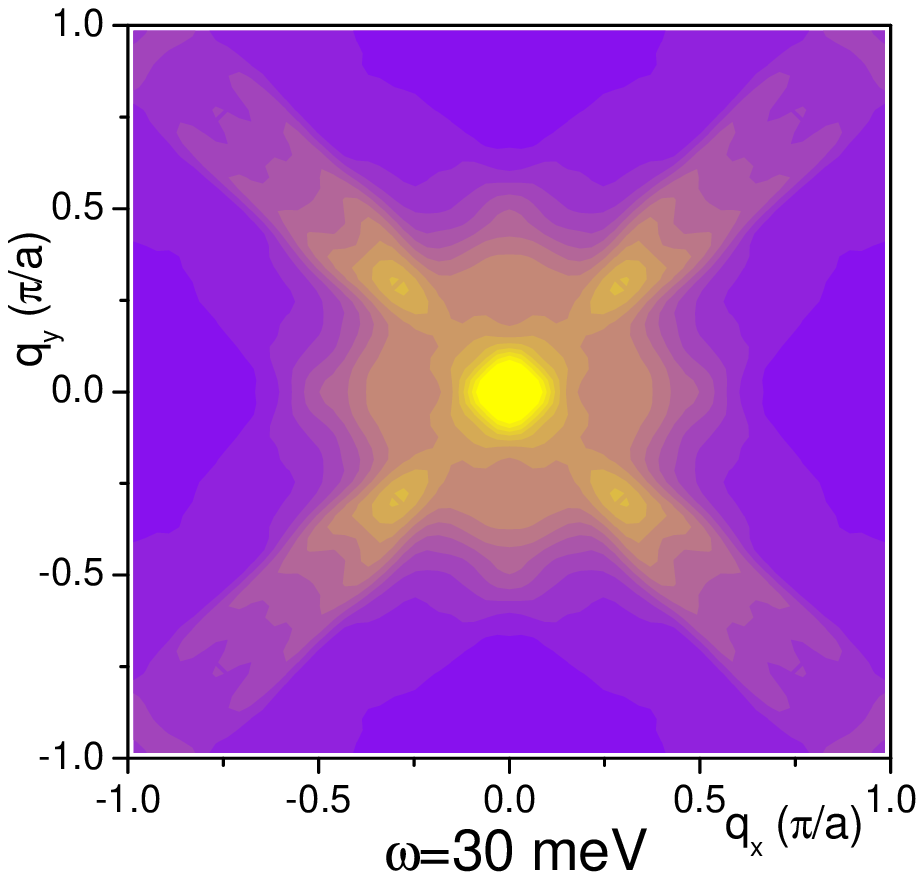}
\caption{Fourier map of LDOS for scattering off a single localized
defect, $\delta \rho_{\rm imp} ({\bf q}, \omega)$ in
(\protect\ref{qs}). We used a localized defect with $u({\bf
q})=1$, $v({{\bf k},{\bf k+q}})=2 \cos(k_x+q_x)-2 \cos(k_y+q_y)$.}
\label{Fig. 4}
\end{center}
\end{figure}

As is clear from Figs.~\ref{Fig. 1} and~\ref{Fig. 2}, the LDOS
also has peaks in its modulation at a number of other wavevectors.
These arise from the two flanking quasiparticle Green's functions
in (\ref{e3}), and are consequently sensitive to all details of
quasiparticle spectrum. These additional features are also present
in the static impurity model (\ref{qs}) whose results are shown in
Fig.~\ref{Fig. 4}
.
It is instructive to compare the static impurity scattering
prediction in Fig~\ref{Fig. 4} to the dynamic spin collective mode
predictions in Figs.~\ref{Fig. 1} and~\ref{Fig. 2}: apart from the
strong contributions at ${\bf q} = \pm 2 {\bf K}_{x,y}$ present in
the latter, the former has the same general features as the large
$\Delta$ cases of the latter. This is one of the main points of
this paper.

The physics of the structure in the static impurity scattering
LDOS modulation, $\delta \rho_{\rm imp} ({\bf q}, \omega)$ in
(\ref{qs}), has already been discussed by Wang and Lee
\cite{dunghai}.
Assuming spatial inversion symmetry, the expression (\ref{qs}) can
be rewritten as:
\begin{eqnarray}
&&\delta \rho_{\rm imp}({\bf q},\omega)=2\int {d^2 k\over
(2\pi)^2}\delta(-\omega^2+\varepsilon_{\bf k}^2+\Delta_{\bf
k}^2)\times\nonumber\\
&&~\left(u({\bf q}){(\omega-\varepsilon_{\bf
k})(\omega-\varepsilon_{{\bf k}+{\bf q}})-\Delta_{\bf
k}\Delta_{{\bf k}+{\bf q}}\over -\omega^2+\varepsilon_{{\bf
k}+{\bf q}}^2+\Delta_{{\bf k}+{\bf q}}^2}-\right.\nonumber\\
&&~\left. v({\bf k},{\bf k+q}){(\omega-\varepsilon_{\bf
k})\Delta_{{\bf k}+{\bf q}}+\Delta_{\bf
k}(\omega-\varepsilon_{{\bf k}+{\bf q}})\over
-\omega^2+\varepsilon_{{\bf k}+{\bf q}}^2+\Delta_{{\bf k}+{\bf
q}}^2}\right).\label{qs1}
\end{eqnarray}
This expression is not a convolution of the product of the density
of states of quasiparticles at energy $\omega$ with momenta
differing by ${\bf q}$, but rather the product of the density of
states of one quasiparticle with the real part of the Green
function of the other. It contains weak square-root divergences at
special nesting wavevectors \cite{doug,vadim}, where constant
energy contours with momenta ${\bf k}$ and ${\bf k}+{\bf q}$
($\omega^2 = \varepsilon_{\bf k}^2+\Delta_{\bf
k}^2=\varepsilon_{\bf k+q}^2+\Delta_{\bf k+q}^2$) are tangent to
each other at one of their the points of contact. Such wavevectors
${\bf q}$ form different lines in the Brillouin zone, but these
are barely visible in our plots. It has been claimed
\cite{dunghai} that the structure at related wavevectors could be
enhanced for strong impurity scattering in a description beyond
the Born approximation. Instead, in Fig~\ref{Fig. 4}, the main
contributions for ${\bf q}$ along the Brillouin zone diagonals
come from values of ${\bf k}$ and ${\bf k} + {\bf q}$ near the
nodal points and the underlying Fermi surface near $\pm(\pi,0),\pm
(0,\pi)$ respectively. We emphasize that both these features of
quasiparticle-scattering induced LDOS modulations also appear in
our model \cite{pvs} of a pinned dynamic spin collective mode. In
particular, as we noted earlier, if $\Delta$ is large, then the
pinned SDW model is equivalent to an effective short-range static
quasiparticle potential, along with additional weak modulations at
$\pm 2 {\bf K}_{x,y}$ (compare Figs~\ref{Fig. 2}, and~\ref{Fig.
4}). For smaller $\Delta$, the effective quasiparticle potential
created by the SDW fluctuations is dynamic, but the qualitative
picture at wavevectors different from $\pm 2 {\bf K}_{x,y}$
remains the same.

In conclusion, we have shown that the model of Ref~\cite{pvs}
displays many of the features observed in recent STM experiments.
Besides the modulations of the LDOS at wavevectors $\pm 2 {\bf
K}_{x,y}$ arising from the spin collective mode, we find other
features related to the spectrum of the fermionic Bogoliubov
quasiparticles (similar physics is also expected from a possible
charge/bond order collective mode). The positions of the maxima at
$\pm 2 {\bf K}_{x,y}$ have a relatively small (but not negligible)
dependence on the tunnelling energy $\omega$, while other features
are more strongly dependent on $\omega$ and the details of the
quasiparticle spectrum . We have also made predictions on the
doping and magnetic field dependence of the relative contributions
of the quasiparticle and spin-collective mode induced modulations.

~\\ {\em Note added:} Some related points and observations appear
in \cite{aharon2}.

\begin{ack}
We thank Seamus Davis, Jung Hoon Han, Aharon Kapitulnik, Steve
Kivelson, Patrick Lee, and Vadim Oganesyan for useful discussions.
A.P. and S.S. were supported by US NSF Grant DMR 0098226. M.V. was
supported by the DFG through SFB 484.
\end{ack}

\end{document}